\documentclass[epj]{svjour}
\usepackage{graphicx}
\usepackage{multirow}
\usepackage{graphics}
\usepackage{hyperref}
\hypersetup{colorlinks = true, allcolors = blue}
\usepackage{authblk}
\usepackage{hepparticles}
\usepackage{hepunits}
\usepackage{hepnames}

\DeclareUnicodeCharacter{2212}{-}

\begin{document}
\title{The W mass and width measurement challenge at FCC-ee}
\author{Paolo Azzurri
}                     
\offprints{}          
\institute{INFN, Sezione di Pisa, Italy}
\date{
{\sl 
(Submitted to EPJ+ special issue:
A future Higgs and Electroweak factory (FCC): Challenges towards discovery, Focus on FCC-ee)}}
%
\abstract{
The FCC-ee physics program will deliver two complementary top-notch precision determinations of the W boson mass, and width. The first and main measurement relies on the rapid rise of the W-pair production cross section near its kinematic threshold.  This method is extremely simple and clean, involving only the selection and counting of events, in all different decay channels.  An optimal threshold-scan strategy with a total integrated luminosity of $12\,{\rm ab}^{-1}$ shared on energy points between 157 and 163\,GeV will provide a statistical uncertainty on the W mass of 0.5\,MeV and on the W width of 1.2\,MeV.  For these measurements, the goal of keeping the impact of systematic uncertainties below the statistical precision will be demanding, but feasible. The second method exploits the W-pair final state reconstruction and kinematic fit, making use of events with either four jets or two jets, one lepton and missing energy.  The projected statistical precision of the second method is similar to the first method's, with uncertainties of $\sim 0.5$ ($1$) MeV for the W mass (width), employing W-pair data collected  at the production threshold and at 240-365\,GeV.  For the kinematic reconstruction method, the final impact of systematic uncertainties is currently less clear, in particular uncertainties connected to the modelling of the W hadronic decays. The use and interplay of Z$\gamma$ and ZZ events, reconstructed and fitted with the same techniques as the WW events, will be important for the extraction of W mass measurements with data at the higher 240 and 365\,GeV energies.
} 
\maketitle

\section{Introduction}
\label{section:intro}

\begin{figure}[htb] 
\includegraphics[width=0.396\textwidth]{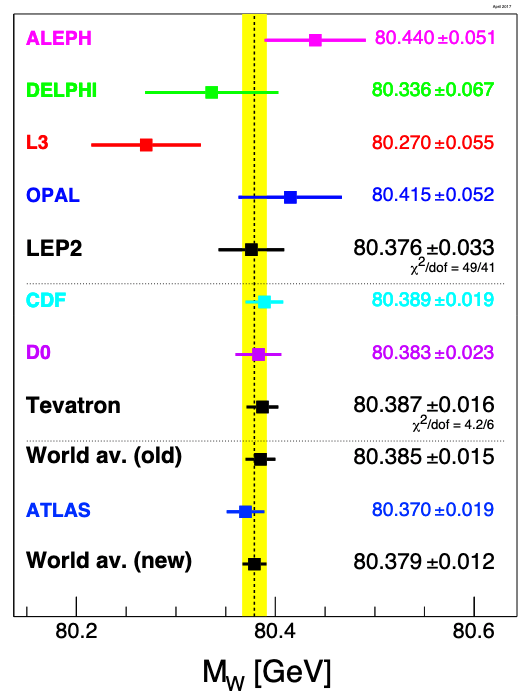}
\includegraphics[width=0.59\textwidth]{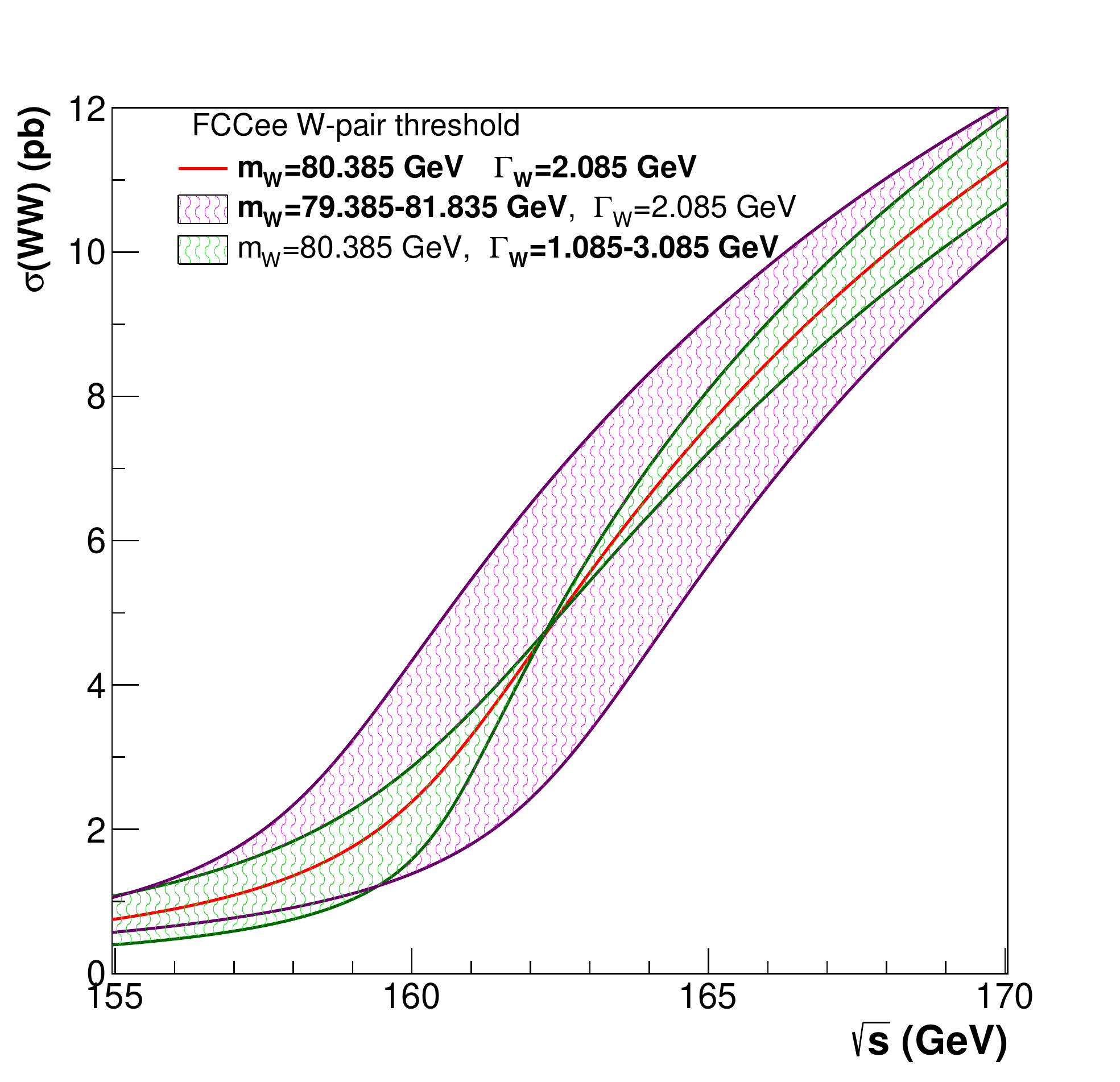}
\caption{(left) Measurements of the W-boson mass by the LEP, Tevatron and LHC experiments\cite{Zyla:2020zbs}. (right) W-pair production cross section as a function of
  the $\rm e^+e^-$ collision energy $E_{\rm CM}$ as evaluated with YFSWW3 1.18~\cite{Jadach:2001uu}.
  The central curve corresponds to the predictions obtained with
  $m_{\rm W}=80.385$~GeV and $\Gamma_{\rm W}=2.085$~GeV.
  Purple and green bands show the cross section curves obtained varying
the W mass and width by $\pm1$~GeV.}
\label{fig:1}       
\end{figure}

The W mass is a fundamental parameter of the standard model (SM)
of particle physics, currently measured with a precision of
12~MeV~\cite{Zyla:2020zbs}, from a combination of LEP, Tevatron and LHC measurements shown in Fig.~\ref{fig:1}.
In the context of precision electroweak tests
the precision of the measurement of the W mass is currently limiting
the sensitivity to possible effects of
new physics~\cite{Baak:2014ora}.

A precise direct determination of the W mass can be achieved
by observing the rapid rise of the W-pair production
cross section near its kinematic threshold.
This method essentially only involves counting events, in all decay channels, and is therefore extremely clean and straightforward. 
In 1996 the LEP collider delivered  $\rm e^+e^-$ collisions at a single
energy point near 161~GeV, with a total integrated luminosity of
about 10 pb$^{-1}$ at each of the four interaction points.
The data was used to measure the W-pair cross section ($\sigma_{\rm WW}$)
at 161~GeV, and extract the W mass with a precision of 200~MeV
\cite{Barate:1997mn,Abreu:1997sn,Acciarri:1997xc,Ackerstaff:1996nk}.
The W mass and width have further  been measured, with better precision,
making use of the full kinematic reconstruction of all decay channels 
at LEP~\cite{Schael:2013ita}, and the partial reconstruction of leptonic 
decays at the Tevatron~\cite{Aaltonen:2013iut} and 
LHC~\cite{Aaboud:2017svj,LHCb:2021bjt} hadron colliders.

Estimates of the W mass and width precision achievable with the FCC-ee physics program are outlined in Ref.~\cite{Abada:2019lih}. Further details and insight are given in the following.

\section{The W-pair cross section lineshape}
The determination of the W mass and width from the W-pair threshold cross section lineshape is presented here. For a basic understanding 
of the statistical and systematic uncertainties, the W mass extraction 
from a single cross section energy point is illustrated first.

Performing a W-pair cross section measurement at a single energy point the statistical sensitivity
to the W mass is given by

\begin{equation}\label{eq:1}
  \Delta m_{\rm W} {\rm (stat)} =
    \left( \frac{d\sigma_{\rm WW}}{dm_{\rm W}}\right)^{-1}
    \frac{\sqrt{\sigma_{\rm WW}}}{\sqrt{\cal L}} \frac{1}{\sqrt{\epsilon p}} = 
    \left( \frac{d\sigma_{\rm WW}}{dm_{\rm W}}\right)^{-1}
    \frac{\sqrt{\sigma_{\rm WW}}}{\sqrt{\epsilon \cal L}} \sqrt{1+\frac{\sigma_B}{\epsilon \sigma_{\rm WW}}} 
\end{equation}

where ${\cal L}$ is the data integrated luminosity, $\epsilon$
the signal event selection efficiency and $p$ the selection purity, 
alternatively expressed in terms of $\sigma_B$, the total selected background
cross section.

A  systematic uncertainty on the background
cross section will propagate to the W mass uncertainty as
\begin{equation}\label{eq:2}
  \Delta m_{\rm W} {(B)} =
    \left( \frac{d\sigma_{\rm WW}}{dm_{\rm W}}\right)^{-1}
    \frac{\Delta \sigma_B}{\epsilon }.
\end{equation}

Other systematic uncertainties as on the acceptance ($\Delta\epsilon$)
and luminosity ($\Delta{\cal L}$) will propagate as 
\begin{equation}\label{eq:3}
  \Delta m_{\rm W} {\rm (A)} = \sigma_{\rm WW}
    \left( \frac{d\sigma_{\rm WW}}{dm_{\rm W}}\right)^{-1}
     \left( \frac{\Delta\epsilon}{\epsilon} \oplus
         \frac{\Delta{\cal L}}{\cal L}\right),
\end{equation}
while theoretical uncertainties on the cross section
($\Delta d\sigma_{\rm WW}$) propagate directly as
\begin{equation}\label{eq:4}
  \Delta m_{\rm W} {\rm (T)} =
  \left( \frac{d\sigma_{\rm WW}}{dm_{\rm W}}\right)^{-1}
  \Delta\sigma_{\rm WW}{\rm (T)}.
\end{equation}

Finally the uncertainty on the center of mass energy $E_{\rm CM}$
will propagate to the W mass uncertainty as
\begin{equation}\label{eq:5}
  \Delta m_{\rm W} {\rm (E)} =
  \left( \frac{d\sigma_{\rm WW}}{dm_{\rm W}}\right)^{-1}
  \left( \frac{d\sigma_{\rm WW}}{dE_{\rm CM}}  \right)
  \Delta E_{\rm CM},
\end{equation}
that can be shown to be limited as 
$\Delta m_{\rm W} {\rm (E)} \leq \Delta E_{\rm CM}/2$,
and in fact for $E_{\rm CM}$ near the threshold
it is $\Delta m_{\rm W} {\rm (E)} \simeq \Delta E_{\rm CM}/2$,
so it is the beam energy uncertainty that propagates directly to
the W mass uncertainty.

In the case of  ${\cal L}=12$~ab$^{-1}$ accumulated
by the FCC-ee data taking in the W-pair threshold energy region,
and assuming an event selection with
$\sigma_B = 300$~fb and $\epsilon = 0.75$,
similar to what was achieved at LEP~\cite{Barate:1997mn},
a statistical precision of $\Delta m_{\rm W}\simeq 0.3$~MeV
is achievable as from Eq.~\ref{eq:1}.
The impact of systematic uncertainties can be kept below the statistical uncertainty
by satisfying the following conditions:
\begin{eqnarray}
  \Delta \sigma_B  < 0.6~{\rm fb} \\
  \left( \frac{\Delta\epsilon}{\epsilon} \oplus \frac{\Delta{\cal L}}{\cal L}\right)
  < 2 \cdot 10^{-4} \\
  \Delta\sigma_{\rm WW}{\rm (T)} < 0.8~{\rm fb} \\
  \Delta E_{\rm CM} < 0.35~{\rm MeV}
\end{eqnarray}
corresponding to precision levels of
$2\cdot 10^{-3}$ on the background,
$2\cdot 10^{-4}$ on acceptance and luminosity,
$2\cdot 10^{-4}$ on the theoretical cross section, and 
$4\cdot 10^{-6}$ on the beam energy.
All of these conditions appear to be challenging yet should be attainable on the
side of experimental systematics, as also discussed later in this essay.
The challenge to
reach the required theoretical precision is discussed in Ref.~\cite{Heinemeyer:2021rgq},
where it is clear that substantial improvements over the current state of the art~\cite{Denner:2005es,Denner:2005fg,Beneke:2007zg,Actis:2008rb}
will be necessary to reach the $2\cdot 10^{-4}$ precision level.

\subsection*{\it W mass and width measurements at two or more energy points}

In the SM the W width is linked to the W mass,
and the Fermi constant, with a $~\sim\alpha_S/\pi$ QCD correction
due to the hadronic decay contributions. The W width is currently measured
to a precision of 42~MeV~\cite{Zyla:2020zbs}.
The first calculations of the W boson width effects
in   $\rm e^+e^- \rightarrow  W^+ W^-$ reactions
have been performed in Ref.~\cite{Muta:1986is},
and revealed the substantial effects of the width on the
cross section lineshape, in particular at energies below the nominal threshold.

From the determination of $\sigma_{\rm WW}$  at a minimum of two
energy points near the kinematic threshold both the W mass and width
can be extracted~\cite{Azzi:2017iih}. 

In the following the YFSWW3 version 1.18~\cite{Jadach:2001uu} program has been
used to calculate  $\sigma_{\rm WW}$ as a function of the
energy ($E_{\rm CM}$), W mass ($m_{\rm W}$) and width ($\Gamma_{\rm W}$).
Figure~\ref{fig:1} shows the W-pair cross section as a function of the
 $\rm e^+e^-$ collision energy with W mass and width values set at the central values
$m_{\rm W}=80.385$~GeV and $\Gamma_{\rm W}=2.085$~GeV,
and with large 1~GeV variation bands around the mass and width 
central values.
It is to be noted that these  does not represent a full state of the art precision on the cross section values, but
deliver a precision that is fully comfortable for all results and conclusions presented
in this paper. In fact the same conclusions in terms of methodology, optimal data taking planning, and projected
precision of the measurements are also reached when making use of the leading order analytical
formulae in Ref.~\cite{Muta:1986is} for the cross section dependencies.

It can be noted that while a variation of the W mass roughly corresponds to a
shift of the cross section lineshape along the energy axis, a variation of the
W width has the effect of changing the slope of the cross section lineshape
rise.
It can also be noted that the W width dependence shows a crossing point
at $E_{\rm CM}\simeq 2m_{\rm W} + 1.5 {\rm GeV }\simeq 162.3$~GeV,
where the cross section is insensitive to the W width. 

\begin{figure}[tbh]
\centerline{
\includegraphics[width=0.50\textwidth]{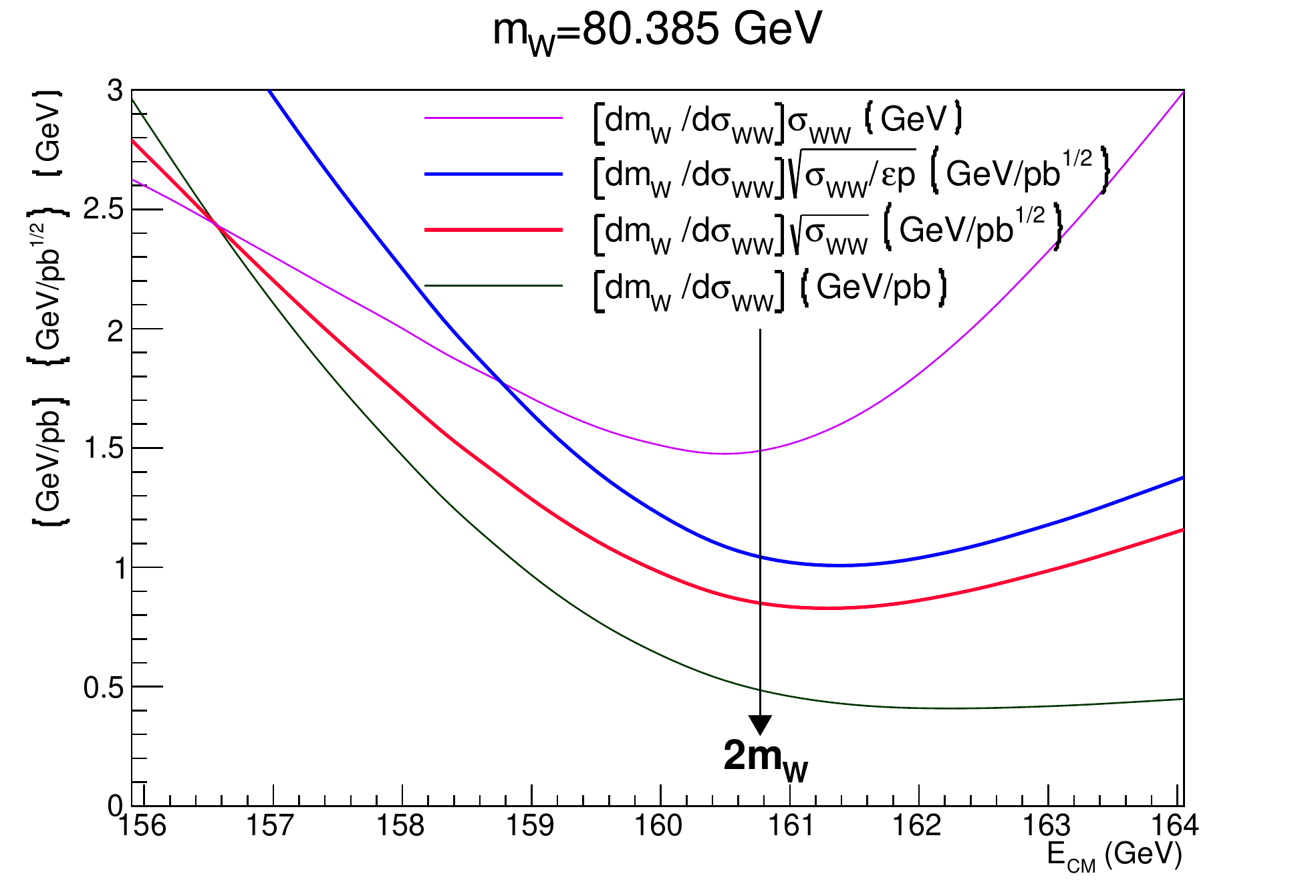}
\includegraphics[width=0.50\textwidth]{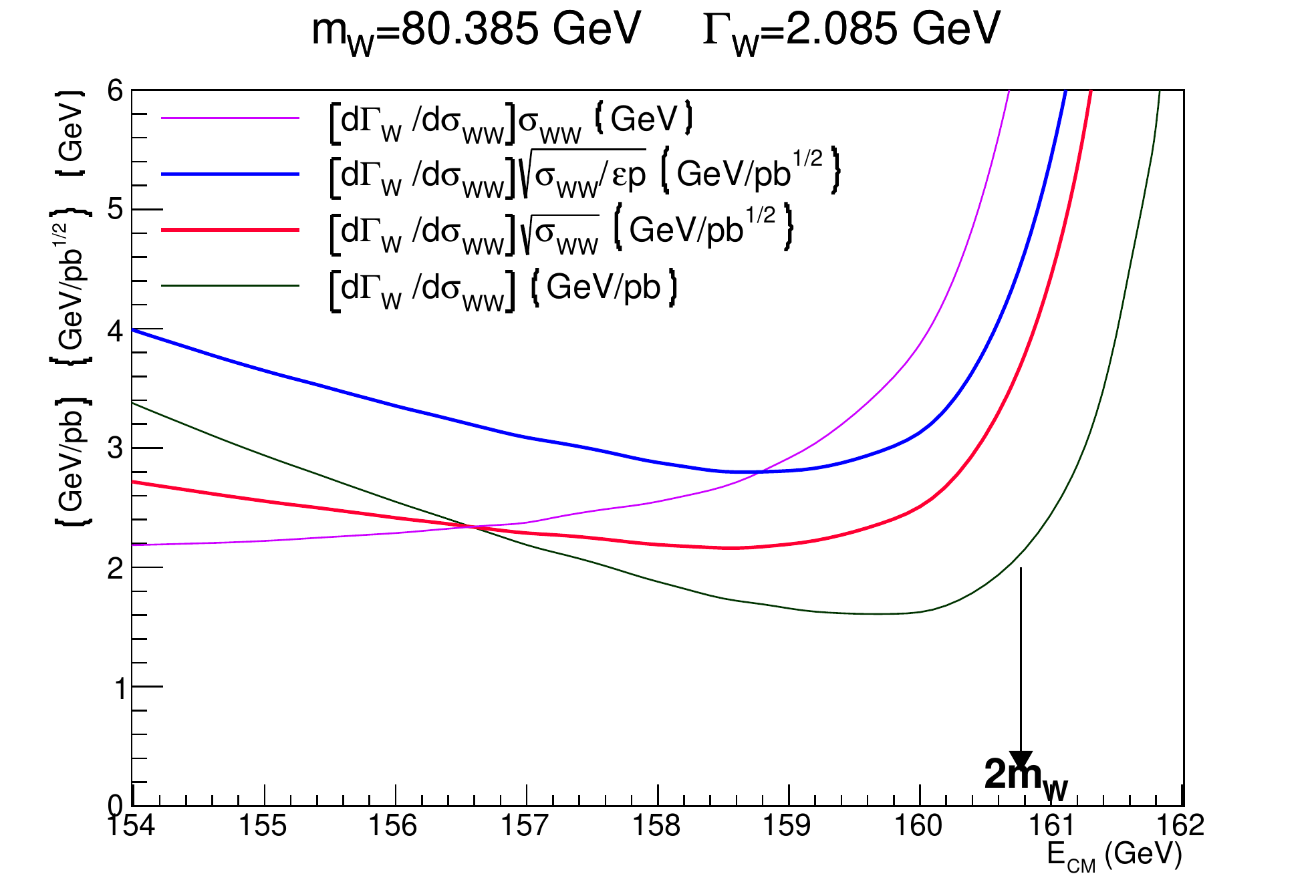}}
\caption{\label{fig:dsww} W-pair cross section differential functions
  with respect to the W mass(left) and width(right), evaluated with YFSWW3 1.18~\cite{Jadach:2001uu}.
 Central mass and width values are set to
  $m_{\rm W}=80.385$~GeV and $\Gamma_{\rm W}=2.085$~GeV.}
\end{figure}

Figure~\ref{fig:dsww} shows the differential functions introduced in
Eq.~\ref{eq:1}~\ref{eq:2}~\ref{eq:3}~\ref{eq:4}, and relevant to the
statistical and systematical uncertainties for a measurement of the W mass and width from the W-pair cross section near the kinematic threshold, similarly as discussed for the single energy point W mass extraction.
For the statistical terms the efficiency and purities are evaluated
assuming an event selection quality with 
$\sigma_B \simeq  300$~fb and $\epsilon \simeq 0.75$.

The minima of the mass differential curves plotted in Fig.~\ref{fig:dsww} left 
indicate the optimal points to take data for a W mass measurement,
in particular minimum statistical uncertainty is achieved
with $E_{\rm CM}\simeq 2m_{\rm W} + 0.6$~GeV$\simeq 161.4$~GeV.

The maximum sensitivity to the W width can be determined from the minima of the curves displayed in
Fig.~\ref{fig:dsww} right.
Note that these curves all diverge at  $E_{\rm CM}\simeq 162.3$~GeV, where
$d\sigma_{\rm WW}/d\Gamma_{\rm W}=0$.
The minima of the width differential curves are spread over a larger $E_{\rm CM}$ area,
with the $\sigma_{\rm WW} ~(d\Gamma_{\rm W}/d\sigma_{\rm WW})$ term decreasing
at lower energies due to the vanishing $\sigma_{\rm WW} $.
This is relevant in the context of an optimal data-taking strategy, if systematic uncertainties
become limiting factors, as discussed later.

If two cross section measurements $\sigma_{1,2}$ are performed at two
energy points $E_{1,2}$, both the W mass and width
can be extracted with a fit to the cross section lineshape.
The uncertainty propagation is given by
\begin{eqnarray}
  \Delta\sigma_1 = \frac{d\sigma_1}{dm} \Delta m + \frac{d\sigma_1}{d\Gamma} \Delta\Gamma
  = a_1 \Delta m + b_1 \Delta\Gamma \\
  \Delta\sigma_2 = \frac{d\sigma_2}{dm} \Delta m + \frac{d\sigma_2}{d\Gamma} \Delta\Gamma
  = a_2 \Delta m + b_2 \Delta\Gamma.
\end{eqnarray}

The resulting uncertainty on the W mass and width is
\begin{equation}
  \Delta m = - \frac{b_2\Delta\sigma_1-b_1\Delta\sigma_2}{a_2b_1 - a_1b_2},
\end{equation}
\begin{equation}
  \Delta \Gamma = \frac{a_2\Delta\sigma_1-a_1\Delta\sigma_2}{a_2b_1 - a_1b_2} .
\end{equation}

If the $\Delta\sigma_{1,2}$ uncertainties on the cross section measurements
are uncorrelated, e.g. only statistical, the linear correlation between
the derived mass and width uncertainties is
\begin{equation}
  r(\Delta m,\Delta \Gamma) = \frac{1}{\Delta m\Delta \Gamma}
  \frac{a_2b_2\Delta\sigma_1^2 + a_1b_1\Delta\sigma_2^2}{(a_2b_1 - a_1b_2)^2}
\end{equation}

\subsection*{\it Optimal data taking configurations}
When planning data taking  at two different
energy points near the W-pair threshold
in order to extract both $m_{\rm W}$ and  $\Gamma_{\rm W}$,
it is useful to figure out which energy points values $E_1$ and $E_2$,
would be optimally suited to obtain the best measurements, also
as a function of the data luminosity fraction $f$ delivered at the 
higher energy point.
For this a full 3-dimensional scan of possible $E_1$, $E_2$ and $f$ values, has been performed,
and the configurations that minimize a given combination
of the expected statistical uncertainties on the mass and the width
$F(\Delta m_{\rm W}, \Delta\Gamma_{\rm W})$ are found.

For example, in order to minimize the simple sum of the statistical uncertainties 
$F(\Delta m_{\rm W}, \Delta\Gamma_{\rm W}) = \Delta m_{\rm W} +\Delta\Gamma_{\rm W}$,
the optimal data taking configuration would be with
\begin{eqnarray}
E_1 = 157.1~{\rm GeV},\quad &   E_2 = 162.3~{\rm GeV},\quad &  f=0.40.
\end{eqnarray}
With this configuration, and assuming a total luminosity of
${\cal L}=12$~ab$^{-1}$, the projected statistical uncertainties would be
\begin{eqnarray}
 \Delta m_{\rm W}=0.5~{\rm MeV} \quad &  {\rm and} \quad\Delta \Gamma_{\rm W}=1.2~{\rm MeV}.
\end{eqnarray}

 Varying the definition of  $F(\Delta m_{\rm W}, \Delta\Gamma_{\rm W})$
used in the optimization does not significantly affect the results. The optimal upper energy is always at
the $\Gamma_{\rm W}$-independent $E_2=162.34$~GeV point, while the optimal lower energy is 
at $(1-2)\Gamma_{\rm W}$ units below the nominal $2m_{\rm W}$ threshold, 
with the precise value depending on the degree to which the definition of $F$ is focused on the W-width measurement.
In a similar way the optimal data fraction to be taken at the
lower off-shell $E_1$ energy point varies according to the
chosen precision targets, with larger fractions more to the benefit
of the W width precision. If a small fraction of data (e.g. $f=$0.05)
is taken off-shell a statistical precision $\Delta m_{\rm W}=0.3$~MeV
is obtainable both with a single- ($ m_{\rm W}$) and the two-parameter
($ m_{\rm W}, \Gamma_{\rm W}$) fit of the lineshape.

Considering that the beam energies $E_b$ that can surely be 
calibrated with resonant
depolarization are such that the spin tune is a half integer, that is 
\begin{equation}
E_b = 0.4406486 ( \nu + 0.5 ) ~{\rm GeV} 
\end{equation}
where $\nu$ is an integer, the scan of energy points can be limited to a grid with $E_{\rm CM}=0.8812972 ( \nu + 0.5 ) ~{\rm GeV}$.
Taking this grid constraint into account the optimal higher energy point for data taking becomes the $E_2=162.62$~GeV for 
$\nu=184$. The corresponding minimum statistical precisions attainable are increased by 5-10\% with respect to the values reported above. For the case of 
minimizing $\Delta m_{\rm W} +\Delta\Gamma_{\rm W}$, 
would be with taking data with
$E_1 = 157.33$~GeV,$E_2 = 162.62$~GeV, $f=0.40$ and yielding 
statistical uncertainties $\Delta m_{\rm W}=0.55$~MeV and 
$\Delta \Gamma_{\rm W}=1.3$~MeV 
assuming a total integrated luminosity ${\cal L}=12$~ab$^{-1}$.

The effects of the beam energy spread effects have also been considered, 
and impact mostly the W width extraction. 
A 10\%-level control on the energy spread  will be sufficient to make the corresponding systematic uncertainties negligible~\cite{Blondel:2019jmp}.

\subsection*{\it Data taking at additional energy points}

In the case of limiting correlated systematics uncertainties,
it can be useful to take data and measure both signal and background cross section at more than two $E_{\rm CM}$ points, in order to reduce background and acceptance uncertainties. 

In particular, for the simultaneous measurement of 
$m_{\rm W}$ and $\Gamma_{\rm W}$ just described, 
taking data at energy points where the differential factors 
$(d\sigma /dm_{\rm W} )^{-1}$, $(d\sigma /d\Gamma_{\rm W})^{-1}$, 
$\sigma(d\sigma /dm_{\rm W} )^{-1}$ and
$\sigma(d\sigma /d\Gamma_{\rm W})^{-1}$, 
are equal, can help cancelling the effect of correlated 
systematic uncertainties of background and acceptance.
Initial investigations in this direction 
have been carried out~\cite{Shen:2018afo}, supporting the presumption
that taking data at more than two energy points improves the 
robustness of the measurement against correlated systematic uncertainties. 

Measuring the W-pair cross section at additional points can also serve to disentangle possible new physics effects,
as for example anomalous triple gauge coupling (TGC) contributions.
The SM-expected steep W-pair cross section rise with energy is proportional to the produced W boson velocity ($\beta_{\rm W}$ )
and is driven by the $t$-channel neutrino exchange process. The contribution of processes with TGCs
follow a  different $\beta_{\rm W}^3$ dependence,  with expected  cancellation effects.
Anomalous TGC contributions would therefore lead to distinctive differences in the W-pair cross section lineshape
also in the threshold region.

\section{W mass and width from the W pair decay kinematics}

In addition to the W mass and width measurements achievable through the W-pair cross sections near the production energy threshold, the W mass and width can also be determined from the kinematic reconstruction of the W-pair decay products. This was the primary method to measure the W mass and width with LEP2 data~\cite{Schael:2013ita}.

In the kinematic reconstruction of the W mass from W-pair decays the fully hadronic ($qqqq$) and semi-leptonic ($qq\ell$\Pnu) final states are exploited, making use of events with either four jets or two jets, one lepton and missing energy. In both cases the reconstructed W mass values are obtained by imposing the constraint that the total four momentum in the event should be equal to the known initial centre-of-mass energy and zero momentum. The four momentum constraints (4C) are implemented by means of a kinematic fit where the measured parameters of the jets and leptons are adjusted, taking account of their measurement uncertainties in such a way as to satisfy the constraints of energy and momentum conservation. The 4C implementation allows to overcome the limitations of jet energy resolution on the W mass reconstruction, and improve the mass resolution from $\sim$10~GeV to $\sim$2~GeV. The kinematic fit of final states with four-momentum conservation constraints can also be applied to other di-boson productions at $E_{\mathrm{CM}}=160-365$ GeV{}, like Z-pairs and Z$\gamma$ events. In the case of Z$\gamma$  final states, also known as radiative returns to the Z-peak, the fit can be shown to lead to a reconstructed Z boson mass as~\cite{Schael:2006mz},

\begin{equation}
 m_{\mathrm{Z}}^2 = s 
 \frac{\beta_1 \sin\theta_1 + \beta_2 \sin\theta_2 - \beta_1 \beta_2  |\sin(\theta_1 + \theta_2)| }{\beta_1 \sin\theta_1 + \beta_2 \sin\theta_2 + \beta_1 \beta_2  |\sin(\theta_1 + \theta_2)| } ,
\label{eq:m12resc}
 \end{equation}
where $\theta_{1,2}$ is the angle of the two leptons or jets from the Z decay, with respect to the photon direction, and $\beta_{1,2}$ are the leptons or jets velocities. The formula in Eq.~\ref{eq:m12resc} is based on fixing the jet directions and velocities to their measured values but rescaling their energies to conserve four-momentum, that follows closely what is done in a kinematic fit.  

Equation~\ref{eq:m12resc} also shows the direct interplay between the reconstructed Z mass and the centre-of-mass energy ($E_{\mathrm{CM}}^2=s$). In practice the Z mass is reconstructed primarily through the decay products direction, and their velocities in the case of hadronic jets, while the energy scale is set by the known collision energy. The same happens with the 4C kinematic reconstruction of W-pairs, where again the energy scale of jets is given by the event $E_{\mathrm{CM}}$ and the angular openings of jets and leptons carry the primary information to determine the W mass, with the jets velocities as the further important ingredient. 

On the other hand, by making use of the value of $m_{\mathrm{Z}}$ precisely measured at the Z pole, the collision energy $E_{\mathrm{CM}}$ can be treated as the parameter to be measured in Eq.~\ref{eq:m12resc}, so that the kinematic fit of radiative decays can be used to determine $E_{\mathrm{CM}}$. This interpretation was used with LEP2 data to cross-check the $E_{\mathrm{CM}}$ values determined by the accelerator~\cite{Schael:2013ita}.

In general a kinematic fit of either Z$\gamma$, ZZ, or WW decays can be equivalently employed either to determine the boson (W or Z) mass assuming a given centre-of-mass energy or, alternatively, the average centre-of-mass energy assuming a fixed boson mass.

\subsection*{\it W-pair reconstruction at FCC-ee data taking energies}

The prospects of the kinematic reconstruction of W-pairs with FCC-ee data
can be estimated taking as a reference existing LEP measurements~\cite{Schael:2006mz}.
In a kinematic reconstruction data analysis 
W-pair decay products are typically forced into four jets using the DURHAM~\cite{Catani:1991hj} algorithm in the hadronic channel,
and into two jets and a lepton in the semi-leptonic channel. 

The reconstructed W mass peak resolution can be remarkably improved  with a four-momentum 
conservation fit (4C) described above,  and eventually
with the additional constraint of equal mass for both W  in each event (5C).
Maximum likelihood template fits of the reconstructed W mass distributions are then used to
extract the value of $m_{\mathrm{W}}$.
With this methodology, used with LEP2 data, it can be estimated that the  combined statistical precision of all FCC-ee data 
would deliver a final precision of around 1~MeV for the W width, and below 0.5~MeV for 
the W mass, matching the precision 
delivered by the threshold cross section lineshape. 

\subsection*{\it Systematic uncertainties}
 
The limitations of systematic uncertainties to the precision of the W mass kinematic reconstruction 
with FCC-ee data are not easy to establish with certainty. 
As for the threshold cross section method, the beam energy uncertainty is reflected directly to the W mass 
reconstruction, in this case through the kinematic fit. 
Beam energy calibration through resonant depolarisation will ensure that this uncertainty will not be a limiting 
factor for the W mass reconstruction with the data taken at 162.6~GeV.  
For the data taken at 240~GeV and 350-365~GeV the analysis 
and kinematic fit of Z$\gamma$ and ZZ events can allow 
to determine the data $E_{\rm CM}$ with high precision, as can be inferred from Eq.~\ref{eq:m12resc} and done with LEP2 data~\cite{Schael:2013ita}.
Extrapolating the LEP2 measurements to the projected FCC-ee data, a statistical precision of around 1~MeV for  $E_{\rm CM}$
would be achievable. This would propagate to a 0.5~MeV systematic uncertainty on the W mass, that matches the projected final
statistical uncertainty, and would therefore not be negligible. 

A number of other systematic uncertainties that were 
relevant for the LEP2 measurements appear to make 
this measurement overall more challenging with respect
to the more simple threshold determinations. 
The most challenging uncertainties are likely to be related 
to non-perturbative QCD modelling of the W-pair decays 
fragmentation, that relates directly to the hadronic jets 
boost, i.e. the $\beta$ factors in Eq.~\ref{eq:m12resc}.
Precise measurements of fragmentation properties of 
Z boson hadronic decays, collected at 
teh Z peak,
will be instrumental to 
build control on the fragmentation properties of weak bosons. 

Finally a simultaneous analysis and kinematic fit of 
WW, ZZ and Z$\gamma$ events, can lead to a determination 
of the $m_{\rm W}/m_{\rm Z}$ ratio where many systematic uncertainties 
common to the three channels can cancel, and the W mass can be 
derived given the independent precision on the Z mass 
($\Delta m_{\rm Z} \simeq 100$~KeV) from the Z peak data. 

\section{Conclusions}\label{section:conclusion}

Among the primary parameters of the standard model, the W mass and 
width are those where a improvement of the experimental determination 
is most desirable. 
These measurements are extremely difficult at high energy hadron colliders,
and the foreseen precision achievable with LHC data is around 10~MeV for the W mass. 

The FCC-ee program will offer the opportunity for a full exploration 
of the W-pair production at the kinematic threshold that will deliver 
a clean and straightforward determination of the W mass and width, with 
respective accuracies of 0.5~MeV and 1.2~MeV.

Complementary and more challenging determinations of the W mass and width 
with FCC-ee data can be obtained through the reconstruction of W-pair 
decay products, from data in the full $E_{\mathrm{CM}}$ range from the production threshold to $E_{\mathrm{CM}}=365$~GeV. 
The projected statistical precision from these other measurements 
are similar to those of the  threshold determinations, but the
impact of systematic uncertainties are more difficult to predict, in particular those arising from the beam energy knowledge and 
the modelling of non-perturbative QCD effects in the W boson hadronic decays.
The ultimate way forward to exploit the kinematic reconstruction method  
could be in the simultaneous analysis of WW, ZZ and Z$\gamma$ events,
making use of the much higher precision of the Z mass from the peak scan, 
and allowing to reduce the impact of  correlated systematic effects.

%
%
%

\bibliographystyle{myutphys}
\bibliography{references}
\end{document}